\documentclass[reprint,prc,amsmath,amssymb,aps,lengthcheck,floatfix,superscriptaddress]{revtex4-1}
\usepackage{balance}
\usepackage{graphicx}
\usepackage{dcolumn}
\usepackage{bm}
\usepackage{hyperref}
\usepackage[utf8]{inputenc}
\usepackage{amsmath}
\usepackage{multirow}
\usepackage{soul}
\usepackage{float}
\usepackage{graphicx}
\usepackage{times}
\usepackage[normalem]{ulem}
\usepackage{color}
\usepackage{cellspace}
\usepackage{amssymb}
\usepackage{xcolor}
\usepackage{lipsum}

\newcommand{\be}{\begin{equation}}
\newcommand{\ee}{\end{equation}}
\newcommand{\bea}{\begin{eqnarray}}
\newcommand{\eea}{\end{eqnarray}}

\newcommand{\comment}[1]{}
\renewcommand\sout{\bgroup \color{red} \ULdepth=-.5ex \ULset}
\usepackage{xcolor}
\graphicspath{ {./images/} }

\def\simge{\mathrel{\rlap{\raise 0.511ex
     \hbox{$>$}}{\lower 0.511ex \hbox{$\sim$}}}}
\def\simle{\mathrel{\rlap{\raise 0.511ex
      \hbox{$<$}}{\lower 0.511ex \hbox{$\sim$}}}}

\usepackage{orcidlink}

\begin{document}
\title{Impact of the Scalar Isovector $\delta$-meson on the description of nuclear matter and neutron star properties}
\author{Lavínia Gabriela Teodoro dos Santos \orcidlink{0009-0003-0202-0515}} \email{lgtds@student.uc.pt}
\affiliation{CFisUC, Department of Physics, University of Coimbra, P-3004 - 516 Coimbra, Portugal}
\author{Tuhin Malik  \orcidlink{0000-0003-2633-5821}} \email{tuhin.malik@uc.pt}
\affiliation{CFisUC, Department of Physics, University of Coimbra, P-3004 - 516 Coimbra, Portugal}
\author{Constança Providência \orcidlink{0000-0001-6464-8023}}
\email{cp@uc.pt}
\affiliation{CFisUC, Department of Physics, University of Coimbra, P-3004 - 516 Coimbra, Portugal}

\date{\today}
\begin{abstract}
The implications of including the scalar isovector $\delta$-meson in a relativistic mean-field description of nuclear matter are discussed. A Bayesian inference approach is used to determine the parameters that define the isovector properties of the model. The properties of nuclear matter and neutron stars are discussed. The inclusion of the $\delta$-meson has only a small effect on the maximum mass of the neutron star (NS) and on the speed of sound in its interior, but it has a strong effect on the radius and the tidal deformability of low and medium mass stars.  This is mainly due to the effect of the $\delta$-meson on the symmetry energy and its slope and curvature at saturation, increasing the range of possible values of these three properties, and in particular allowing positive values of the symmetry energy curvature. Due to the effect of the $\delta$-meson on the symmetry energy, the proton content of the star is also strongly affected. The inclusion of the $\delta$-meson in the relativistic mean-field description of nuclear matter extends the phase space spanned by the model, allowing for a more flexible density dependence of the symmetry energy compatible with experimental, observational, and ab initio constraints.
\end{abstract}
\maketitle

\section{Introduction}
In \cite{PhysRevC.65.045201}, the authors have studied the effects of introducing the $\delta$ isovector scalar meson into a relativistic mean field (RMF) description of nuclear matter, alongside the usual scalar-isoscalar $\sigma$ meson and the vector-isoscalar $\omega$ meson and the isovector $\rho$ meson \cite{Serot1984}. The idea was to have an isovector channel with a structure similar to the isoscalar one, i.e. for each channel a scalar and a vector meson is introduced.  They have shown that this extra meson has a strong effect on the symmetry energy, in particular on its slope and curvature. Another effect is the splitting of the neutron and proton masses. These effects were further discussed in \cite{Gaitanos:2003zg}, where within an RMF with density dependent nucleon-meson couplings, the $\delta$ meson was also included.  Again, a strong effect on the symmetry energy at high densities and on the neutron/proton effective mass splitting was verified. The authors have also shown that the introduction of the $\delta$-mesons gives important contributions to heavy ion reactions. In \cite{Menezes:2004vr} the authors have studied the effect of the $\delta$ meson on the properties of NS. {They have shown that its inclusion makes the EOS slightly harder and gives larger NS radii and slightly larger maximum masses when only nucleons are considered.} However, the inclusion of hyperons leads to a greater softening, with smaller radii and maximum masses. Another effect was a reduction of the neutrino fraction in proto-neutron stars with a trapped neutrinos. Some of these effects have also been discussed in \cite{Klahn:2006ir}. The effect of the $\delta$-meson on the dynamical spinodal was discussed in \cite{Pais:2009rr}. The model DDME$\delta$ including the $\delta$-meson was quite successfully constrained to nuclear properties and {\it ab-initio} non-relativistic and relativistic Brueckner calculations of symmetric and asymmetric nuclear matter \cite{Roca-Maza:2011alv}. Recently, several papers have further discussed the inclusion of the $\delta$ meson in an RMF description of nuclear matter, see, among others, \cite{Miyatsu:2022wuy,Li:2022okx,Lopes:2023dnx,Reed:2023cap}. In \cite{Miyatsu:2022wuy,Li:2022okx} the authors introduce a term that mixes the two scalar mesons and softens the symmetry energy.  In \cite{Reed:2023cap} the authors propose to reconcile the results of PREX2 and CREX by introducing the $\delta$ meson. It is shown that a possible description of both experimental data at 1$\sigma$ is possible with parameterizations that give a very large curvature of the symmetry energy. Correlations between the symmetry energy slope and symmetry energy curvature or the radius of a 1.4$M_\odot$ star are examined.

Neutron stars (NS) have been widely studied because they are composed of very asymmetric matter that is not possible to study in the laboratory. The determination of the high density of asymmetric nuclear matter is still under investigation, however, it is expected that measurement of the mass of massive stars as the pulsars PSR J1614-2230,  PSR J0348+0432 and PSR J0740+6620 \cite{Demorest2010,Antoniadis2013,Fonseca2016,NANOGrav:2019jur,Fonseca:2021wxt},  the detection of gravitational wave signals from two merging NS (such as GW170817 or GW190425) \cite{LIGOScientific:2017ync,LIGOScientific:2020zkf} and further detections by the LIGO Virgo collaboration and measurements of the mass and radius of pulsars of pulsars PSR J0030+0451, PSR J0740+6620 and PSR J0437-4715 by  the  NASA's Neutron Star Interior Composition Explorer (NICER)  X-ray telescope \cite{Riley:2019yda, Miller:2019cac,Riley:2021pdl, Miller:2021qha, Raaijmakers:2021uju,Watts:2024ozl} allow the inclusion of new observational constraints in the equation of state (EOS) evaluation. In addition to observational data, also nuclear saturation properties and neutron matter properties in \textit{ab-initio} calculations below the saturation density ($\rho_{0}$).

In the present study, we aim at calibrating the EOS for $\beta$-equilibrium matter constrained by nuclear matter properties within a RMF description of nuclear matter that includes the $\delta$-meson. The term mixing the $\omega\rho$ mesons is also considered but not a term mixing the $\delta\sigma$ mesons. Three calibrated and unified  EOS proposed in \cite{Malik:2024nva} having different maximum masses. The inner crust of these EOS has been obtained within a compressible liquid drop model from the same Lagrangian density that describes the core EOS.  The isovector channel determined by the $\delta$ and the $\rho$-mesons  is allowed to vary for each EOS and it is  constrained within a Bayesian inference calculation. This means that three couplings, the Yukawa coupling of the $\rho$ and $\delta$-mesons to the nucleon and the coupling of the mix $\omega\rho$ term are taken as free parameters to be determined from the Bayesian calculation. The parameters that determine the behavior of the symmetric nuclear matter will be kept fixed and equal to the one of the three underlying EOS. This will allow us to discuss the role of the isovector channel on the NS properties.   The effect of the introduction of the $\delta$-meson is discussed, in particular, the behavior of the symmetry energy and the NS properties. It was shown in \cite{Vidana2009}  from a large set of relativistic and non-relativistic nuclear models that the slope and the curvature of the symmetry energy could be linearly anti-correlated. A similar conclusion was drawn in \cite{Char:2023fue}.  However, this correlation was not obtained in  other studies performed within a Bayesian framework \cite{Malik:2022zol,Malik:2023mnx,Salinas:2023nci}.  A strong linear correlation between the radius of a 1.4 $M_\odot$ star and the slope $L$ of the symmetry energy was found in \cite{Salinas:2023nci}, contrary to the weak correlation determined in \cite{Char:2023fue,Scurto:2024ekq}. A correlation similar to this one was obtained in \cite{Alam:2016cli} but only for low mass NS. The particularity of the study undertaken in \cite{Salinas:2023nci} was that the bulk properties of symmetric nuclear matter at saturation density were almost unchanged during the Bayesian inference process. 

The paper is organized as follows: in Sec. \ref{sec2} we review the model; in Sec. \ref{sec3} the Bayesian inference methodology adopted is explained; results are presented and discussed in Sec. \ref{sec4}, and some conclusions are drawn in the last section.

\section{Formalism}\label{sec2}
In a relativistic mean-field approach, it is possible to model nuclear matter through a Lagrangian density involving baryons and mesons, the latter creating mean fields through which nucleons interact among them. Generally, nuclear matter is described through the introduction of three mesons, the scalar-isoscalar $\sigma$-meson, the vector-isoscalar $\omega$-meson and the vector-isovector $\rho$-meson. In the present study we will also introduce the scalar-isovector $\delta$-meson. The  $\sigma$ and $\omega$ mesons are  responsible, respectively, for the long-distance attraction and short-distance repulsion between nucleons,  the  $\delta$ and $\rho$ mesons, both isovectors and, therefore, distinguishing protons from neutrons (nucleons), in particular, the first one the  nucleons’ effective masses and the second the difference in energy when the constitution of matter is not symmetric. The Lagrangian is then written as
\begin{equation}
    \begin{split}
    \mathcal{L} &= \bar{\Psi}[i\gamma_{\mu}\partial^{\mu}-(M_{N}-g_{\sigma}\phi-g_{\delta}\boldsymbol{{\tau}}\cdot\boldsymbol{{\delta}})-g_{\omega}\gamma_{\mu}\omega^{\mu} \\
    &-\frac{g_{\rho}}{2}\gamma^{\mu}\boldsymbol{{\tau}}\cdot\boldsymbol{b}_{\mu}]\Psi+\frac{1}{2}(\partial_{\mu}\phi\partial^{\mu}\phi-m_{\sigma}^{2}\phi^{2})-U(\phi) \\
    &+\frac{1}{2}m_{\omega}^{2}\omega_{\mu}\omega^{\mu}+\frac{1}{2}m_{\rho}^{2}\boldsymbol{b}_{\mu}\cdot\boldsymbol{b}^{\mu}+\frac{1}{2}(\partial_{\mu}\boldsymbol{{\delta}}\cdot\partial^{\mu}\boldsymbol{{\delta}}-m_{\delta}^{2}\boldsymbol{{\delta}}^{2}) \\
    &-\frac{1}{4}\boldsymbol{G}_{\mu\nu}\boldsymbol{G}^{\mu\nu}
   +\frac{\xi}{4!}g_{\omega}^4(\omega_{\mu}\omega^{\mu})^2 
   +g_{\omega\rho}g_{\rho}^{2}\boldsymbol{b}_{\mu} \cdot \boldsymbol{b}^{\mu} g_{\omega}^{2}\omega_{\mu}\omega^{\mu},
    \end{split}   
    \label{eq:lagrangian}
\end{equation}
where $\phi$ is the $\sigma$-meson field, $\omega_{\mu}$ the $\omega$-meson field, $\boldsymbol{b}_{\mu}$ the charged $\rho$-meson field, and $\boldsymbol{\delta}$ the isovector scalar field of the $\delta$-meson. The tensors are defined as $F_{\mu\nu}\equiv\partial_{\mu}\omega_{\nu}-\partial_{\nu}\omega_{\mu}$ and  $\boldsymbol{G}_{\mu\nu}\equiv\partial_{\mu}\boldsymbol{b}_{\nu}-\partial_{\nu}\boldsymbol{b}_{\mu}-g_\rho \boldsymbol{b}_{\mu}\times \boldsymbol{b}_{\nu}$. $U(\phi)$ is the nonlinear potential of the $\sigma$ meson: $U(\phi)=\frac{1}{3}b\, M_N\,{(g_\sigma \phi)}^{3}+\frac{1}{4}c{(g_\sigma\phi)}^{4}$ \cite{PhysRevC.65.045201}. We also include a non-linear $\omega$ self-interaction $\omega^4$ responsible for the softening of the EOS at large densities, and the $\omega\rho$  mixing terms important to define the density dependence of the symmetry energy. 

In the present study we analyze the effect of the inclusion of the $\delta$-meson on the isovector channel of the EOS. This is done by considering three representative EOS that do not include the $\delta$-meson. Through a Bayesian inference procedure, the couplings that define the isovector behavior of the model, i.e. $g_\rho$, $g_\delta$ and $g_{\omega\rho}$, are determined by imposing a set of nuclear matter and observational constraints. The three EOS were chosen from a dataset of previous work \cite{Malik:2023mnx,Malik:2024nva}, obtained from a Lagrangian density without the $\delta$-meson (Eq. \ref{eq:lagrangian}). This choice was based on the stiffness of the EOS,  more specifically on the maximum NS mass, and  a "soft", a "moderate" and a "stiff" EOS were chosen (EOS8, 20 and 21, respectively, of ref. \cite{Malik:2024nva}). The parameters for the original EOS before including the $\delta$-meson are shown on Table \ref{tab:params_props}, as well as some nuclear matter  properties.

\begin{table*}[]
\centering
\caption{The parameters for the 3 models chosen from previous works (\cite{Malik:2023mnx} and \cite{Malik:2024nva}). Specifically, B and C are $b \times 10^3$ and $c \times 10^3$, respectively. The nucleon, $\omega$-meson, $\sigma$-meson, and $\rho$-meson masses considered are 939, 782.5, 500, and 763 MeV, respectively. Along with the median and associated 90\% CI of some properties for the 3 models.}
\label{tab:params_props}
\begin{tabular}{ccccccccc}
\hline \hline 
\textbf{EOS}                         & $g_{\sigma}$       & $g_{\omega}$        & $g_{\rho}$        & B                       & C                         & $\xi$        & \multicolumn{2}{c}{$g_{\omega\rho}$}                                \\ \hline
EOS8                                & 8.637377 & 10.348224 & 11.228904 & 3.910898                & -2.158740                 & 0.001478 & \multicolumn{2}{c}{0.078386}                          \\ 
EOS20                               & 9.554944 & 11.640795 & 14.692091 & 3.887915                & -4.661381                 & 0.003635 & \multicolumn{2}{c}{0.043001}                          \\ 
EOS21                               & 9.608190 & 11.957725 & 12.191950 & 3.117923                & -4.098400                 & 0.000255 & \multicolumn{2}{c}{0.058744}                          \\ \hline
\multirow{2}{*}{\textbf{Properties}} & $E_{\text{sym}}$     & L         & $K_{\text{sym}}$      & $R_{1.4}$                    & $\Lambda_{1.4}$ & $M_{\text{max}}$     & $\rho_c(M_{\text{max}})$ & $K_0$                        \\ 
                                     & (MeV)    & (MeV)     & (MeV)     & (km)                   &                            & ($M_{\odot}$)     & (fm$^{-3}$)                    & (MeV)                     \\ \hline
EOS8                                & 28.5     & 35.6      & -77       & 12.59                   & 511                       & 2.24     & 0.955                     & 268                       \\ 
EOS20                               & 33.8     & 32.2      & -18       & 12.91                   & 514                       & 2.38     & 0.885                     & 200                       \\ 
EOS21                               & 29.1     & 42.5      & 56        & 12.95                   & 638                       & 2.57     & 0.791                     & 217                       \\ \hline
\multirow{2}{*}{\textbf{Properties}} & $R_{2.0}$     & $R_{2.08}$     & $R_{\text{max}}$      & $c_s^2$ & $x_p$                        & $M_{\text{dUrca}}$   & dUrca density             & $Q_0$                        \\ 
                                     & (km)     & (km)      & (km)      & ($c^2$)                   & ($M_{\odot}$)                      & (fm$^{-3}$)   & (fm$^{-3}$)                    & \multicolumn{1}{l}{(MeV)} \\ \hline
EOS8                                & 12.34    & 12.17     & 11.23     & 0.658                   & 0.117                     & ...      & 2.15                      & \multicolumn{1}{l}{-333}  \\ 
EOS20                               & 12.57    & 12.66     & 11.70     & 0.597                   & 0.146                     & 1.94     & 0.56                      & \multicolumn{1}{l}{-567}  \\ 
EOS21                               & 13.17    & 13.14     & 12.13     & 0.767                   & 0.138                     & 2.08     & 0.55                      & \multicolumn{1}{l}{-337}  \\ \hline
\end{tabular}
\end{table*}

\section{Bayesian Inference}\label{sec3}
Bayesian inference stands as a powerful statistical method for parameter estimation and forming predictions by integrating prior knowledge with observed data. Unlike frequentist approaches, Bayesian methods incorporate data uncertainty for estimating both parameters and their distribution. This process is systematically described by Bayes' theorem, it is represented as \cite{Wesolowski:2015fqa,Furnstahl:2015rha,Ashton2019,Landry:2020vaw}:
\begin{equation}
    P(\theta \mid X) = \frac{P(X \mid \theta) \cdot P(\theta)}{P(X)}
\end{equation}
where:
\begin{itemize}
    \item $P(\theta \mid X)$ is the posterior probability of the parameter $\theta$ after observing the evidence $X$.
    \item $P(X \mid \theta)$ is the likelihood, which is the probability of observing the evidence given the parameter $\theta$.
    \item $P(\theta)$ is the prior probability of the parameter $\theta$, which represents our beliefs about $\theta$ before observing the evidence.
    \item $P(X)$ is the evidence or marginal likelihood, which is a normalizing constant ensuring that the posterior probabilities sum to 1.
\end{itemize}
In Bayesian inference, the initial belief $P(\theta)$ is revised as new data $X$ are observed, resulting in the posterior belief $P(\theta \mid X)$. This updating process allows for continuous refinement of knowledge, as initial beliefs are constantly modified by incoming evidence.

We have considered the minimal data set in the likelihood, which includes several fundamental empirical nuclear saturation properties (NMP). These properties consist of the binding energy $\epsilon_0$ for symmetric nuclear matter, the incompressibility of nuclear matter $K_0$, and the symmetry energy $E_{\rm sym}$ at saturation density $\rho_0$. In addition, we have included state-of-the-art theoretical constraints, such as the equation for pure neutron matter derived from $\chi$EFT at low densities (see table \ref{tab:constraints}). 

\begin{table}[!t]
\centering
\caption{Constraints applied to the dataset within the Bayesian inference: energy per nucleon for symmetric nuclear matter (SNM) $\epsilon_0$, incompressibility modulus $K_0$, and symmetry energy $E_{\rm sym,0}$, at the saturation density $\rho_0$; pressures for pure neutron matter (PNM) measured at baryon densities of 0.08, 0.12, and 0.16~fm$^{-3}$ from $\chi$EFT \cite{Hebeler2013}.}
    \label{tab:constraints}
     \setlength{\tabcolsep}{8.2pt}
        \renewcommand{\arraystretch}{1.4}
\begin{tabular}{cccc}
\hline \hline 
\multicolumn{2}{c}{Quantity}                      & Constraint                                    & Ref.                     \\ \hline
               {NMP} & $\rho_0$                & $0.153\pm0.005$ MeV                       & \cite{Typel1999}                 \\
                     & $\epsilon_0$            & $-16.1\pm0.2$ MeV                         & \cite{Dutra:2014qga}                 \\
                     & $K_0$                   & $230\pm40$ MeV                            & \cite{Shlomo2006,Todd-Rutel2005}  \\
                     & ${E}_{\rm sym,0}$       & $32.5\pm1.8$ MeV                          & \cite{Essick:2021ezp}              \\
 \\                    
               {PNM} & $P_1^{\rm PNM}$     & $0.521\pm0.091$ MeV fm$^{-3}$                 &  \cite{Hebeler2013} \\
                     & $P_2^{\rm PNM}$     & $1.262\pm0.295$ MeV fm$^{-3}$                 &   \cite{Hebeler2013} \\
                     & $P_3^{\rm PNM}$     & $2.513\pm0.675$ MeV fm$^{-3}$                 &   \cite{Hebeler2013}   \\
                      \hline
\end{tabular}
\end{table}

The likelihood functions employed for the different quantities investigated in this article are described below. For nuclear saturation properties (NMP), we use a Gaussian likelihood, given by
    \begin{equation}
    \mathcal{L}^{\rm NMP}(D |\boldsymbol{\theta})=\prod_j \frac{1}{\sqrt{2 \pi \sigma_j^2}} e^{-\frac{1}{2}\left(\frac{d_j-m_j(\boldsymbol{\theta})}{\sigma_j}\right)^2} , \label{eq:gausslikelihood}
    \end{equation}
 where the index $j$ runs over all the data points (see table \ref{tab:constraints}), $d_j$ are the constraining data values, $m_j(\boldsymbol{\theta})$ are the model values corresponding to the set of model parameters $\boldsymbol{ \theta}$, and the $\sigma_j$ are the adopted uncertainties of the data.
The $\chi$EFT constraints to the pure neutron matter (PNM) were enforced using a super-Gaussian box function probability with a minor tail, expressed as,
\begin{align}
    \mathcal{L}^{\rm PNM}(D |\boldsymbol{\theta}) =\prod_j \frac{1}{2 \sigma_j^2}\frac{1} {\exp{\left(\frac{|d_j-m_j(\boldsymbol{\theta})|-\sigma_j}{0.015}\right)}+1} , \label{eq:boxlikelihood}
\end{align}
where $d_j$ is the median value and $\sigma_j$ represents two times the uncertainty of the $j^{\rm th}$ data point from the $\chi$EFT constraints \cite{Hebeler2013}.
The total likelihood in the inference analysis is given by
\begin{equation}
    \mathcal{L}^{\rm total}=\mathcal{L}^{\rm NMP} \times \mathcal{L}^{\rm PNM}.
\end{equation}
In this study, we have employed a nested sampling method utilizing the \texttt{PyMultiNest} \cite{Buchner:2014nha} algorithm with 2000 live points. This approach has yielded around 6000 samples in the final posterior, indicating a good posterior quality.

\section{Results and discussion}\label{sec4}
{We have selected three different parameterizations within the relativistic mean field (RMF) framework as base models, to describe the equation of state (EOS) without the inclusion of the $\delta$-meson, specifically EOS8, EOS20, and EOS21, as presented in Ref. \cite{Malik:2024nva}. These parameterizations were chosen to provide a reasonable spread in terms of the predicted maximum mass of neutron stars (NS). The maximum masses for these EOS models are 2.2 M$_\odot$, 2.3 M$_\odot$, and 2.5 M$_\odot$, respectively.}

\begin{figure*}[t]
    \centering
    \includegraphics[width=1.\linewidth]{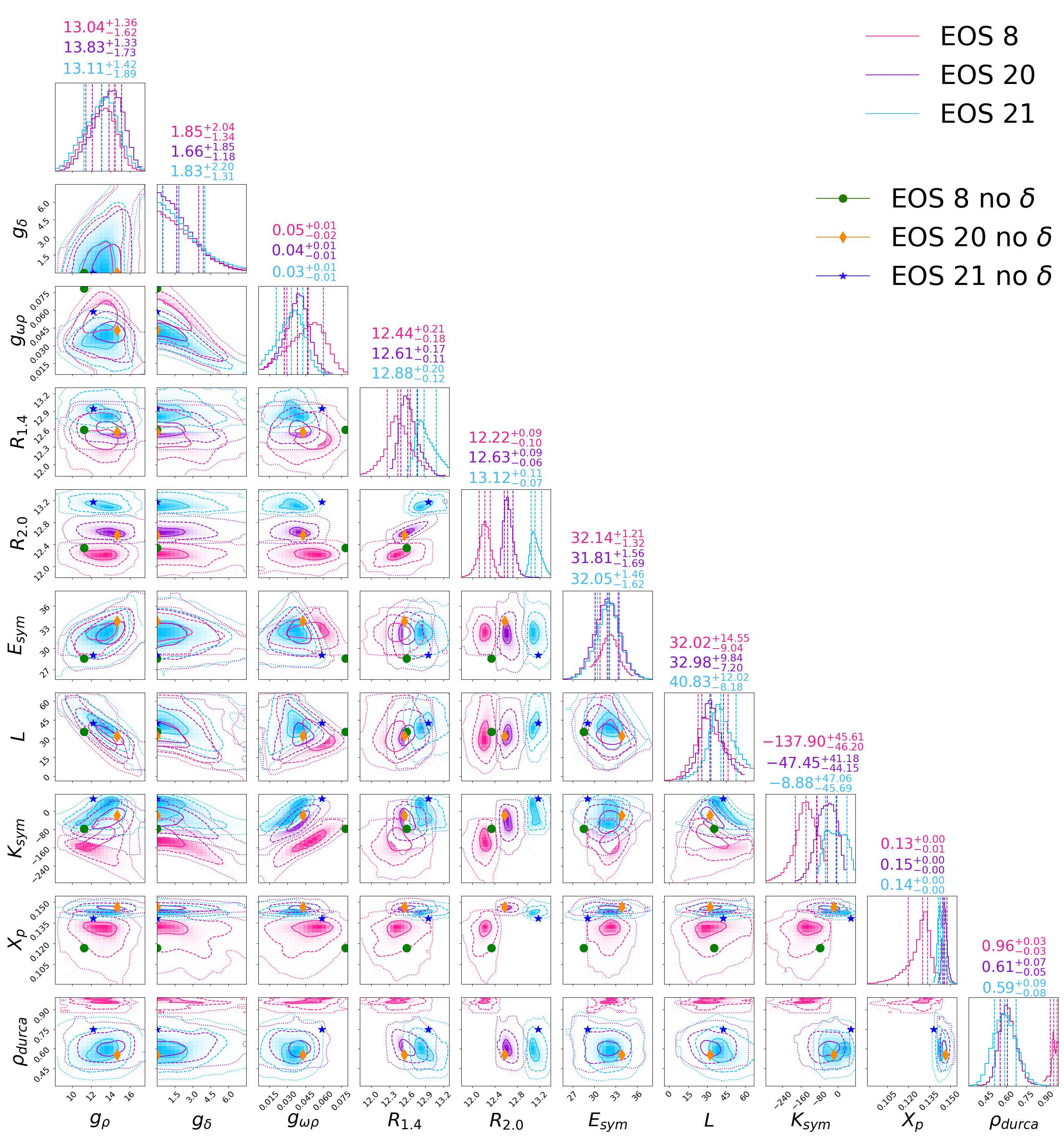}
    \caption{Corner plot \cite{corner} displaying the model parameters \(g_\rho\), \(g_\delta\), and \(g_{\omega\rho}\); neutron star radii for masses of 1.4 and 2 \(M_\odot\); nuclear matter properties, including the symmetry energy \(E_\text{sym}\equiv {\cal S}(\rho_0)\), the slope of the symmetry energy \(L\), and its curvature \(K_\text{sym}\) at saturation density; the proton fraction \(X_p\) at the maximum mass \(M_\text{max}\); and the dUrca onset density \(\rho_\text{dUrca}\) for the obtained posterior distributions of EOS sets 8 (pink), 20 (purple), and 21 (light blue). Each diagonal subplot includes three curves per color, representing the \(1\sigma\) (dotted), \(2\sigma\) (dashed), and \(3\sigma\) (solid) confidence intervals (CIs). The mean value along with the \(1\sigma\) confidence interval is shown above in the diagonal one-dimensional plots. The circular, diamond, and star markers in each respective off-diagonal 2D distribution represent the corresponding values from the base model without the inclusion of the \(\delta\)-meson.}
    \label{fig:corner-NS}
\end{figure*}

\begin{figure*}[t]
    \centering
    \includegraphics[width=1\linewidth]{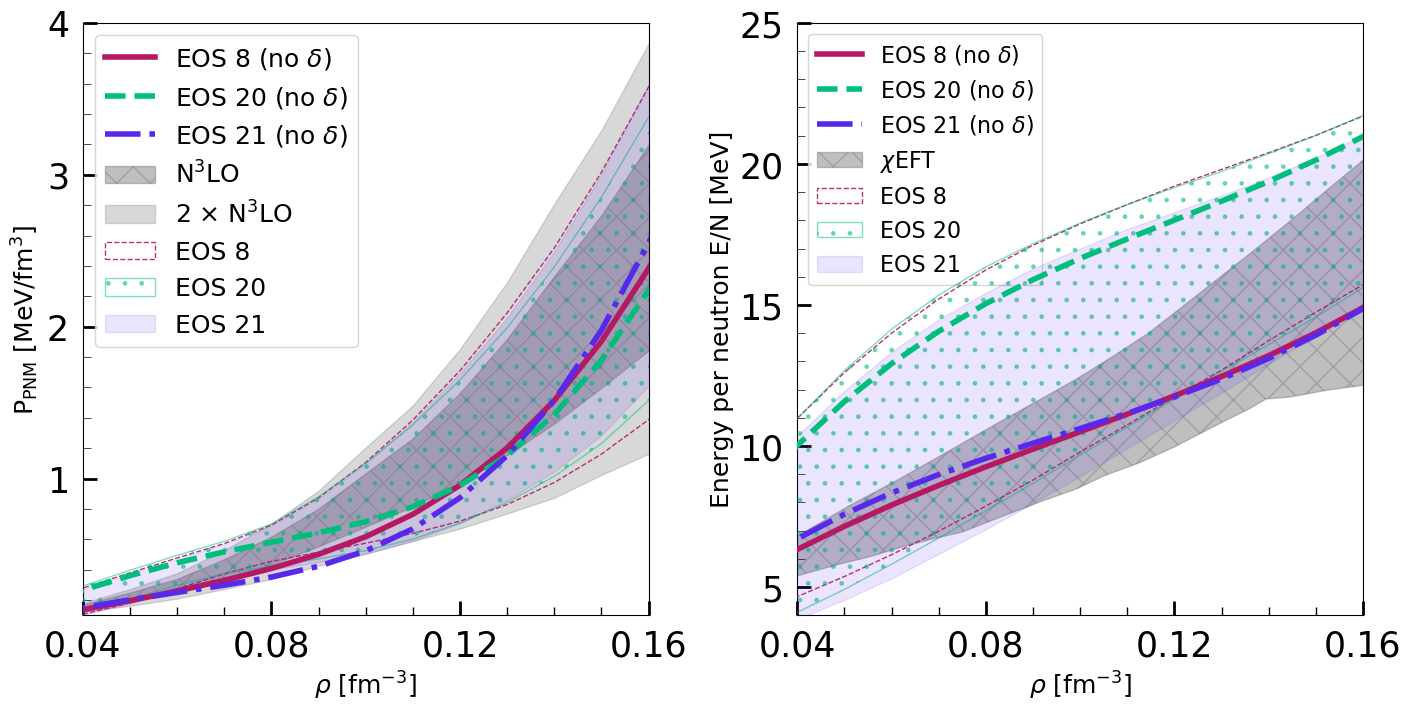}
    \caption{{We present the 90\% credible interval (CI) regions for pure neutron matter (PNM)  pressure (left panel) and energy per neutron of PNM (right panel) as functions of number density for the EOS8, EOS20, and EOS21 posteriors with the inclusion of the $\delta$-meson. The corresponding results for these base EOS models without the $\delta$-meson are also shown by thick lines. In the left panel, the pressure is compared to the $\chi$EFT N3LO results for pure neutron matter from Hebeler et al. (2013) \cite{Hebeler2013}, while in the right panel, the energy per neutron is compared to various $\chi$EFT calculations compiled in Ref. \cite{Huth:2021bsp}.}}
    \label{fig:pnm_chieft}
\end{figure*}

{We have generalized each of these three EOS parameterizations by including the $\delta$-meson, following the formalism outlined in Section \ref{sec2}. In this process, we have adjusted three key parameters: $g_\rho$ (the coupling constant for the $\rho$-meson), $g_\delta$ (the coupling constant for the $\delta$-meson), and the couplings product $g_{\omega\rho} \times g_\rho^2$ (representing the product of the  non-linear coupling between the $\omega$ and $\rho$ mesons and the square of the $\rho$-meson coupling), so that the third parameter is of the same order of magnitude of the other two. {Within the Bayesian inference framework, we conducted sampling of these three parameters, integrating the set of empirical and theoretical constraints outlined in Sec. \ref{sec3} (refer to table \ref{tab:constraints}).} This approach allowed us to explore the impact of the $\delta$-meson on the neutron star EOS and the corresponding uncertainty in a systematic manner. In what follows, we examine the characteristics of the EOS that emerge from the inference process.}

\begin{figure*}[t]
    \centering
    \begin{tabular}{cc}
    \includegraphics[width=0.50\linewidth]{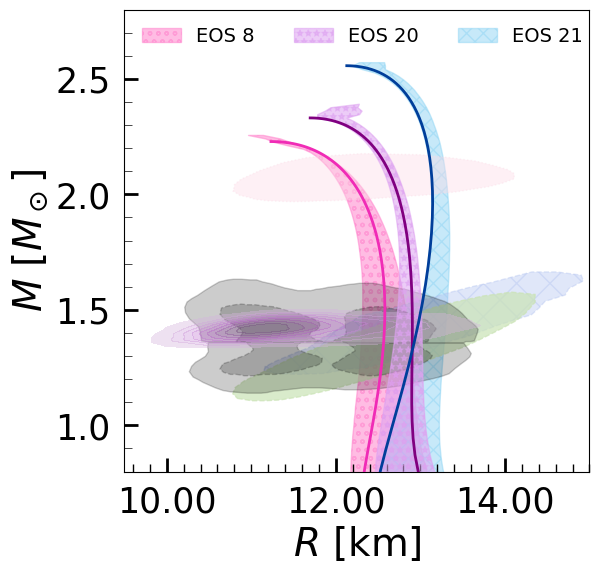}&
    \includegraphics[width=0.50\linewidth]{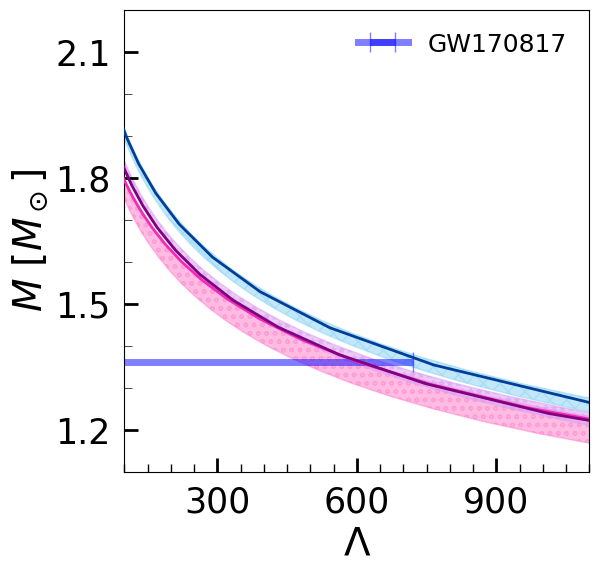}
    \end{tabular}
    \caption{The 90\% CI mass versus radius (MR) and versus $\Lambda$ (M$\Lambda$) distributions obtained with the models that include the $\delta$-meson, together with the mass radius curves (full lines) obtained with the original EOS (EOS8, EOS20 and EOS21) with no $\delta$-meson. The gray regions indicate the 90\% (light) and 50\% (dark) CI constraints from the binary components of GW170817 \cite{LIGOScientific:2018cki}. The mass-radius posterior for the latest PSR J0437-4715 (orchid purple), the 1$\sigma$ (68\%) CI for the 2D posterior distribution in the mass-radii domain for the millisecond pulsar PSR J0030 + 0451 (pale green and pastel blue) \cite{Riley:2019yda,Miller:2019cac}, as well as the PSR J0740 + 6620 (light rose pink) \cite{Riley:2021pdl,Miller:2021qha} from the NICER X-ray data are also shown. The blue bars in the right panel represent the tidal deformability at 1.36 $M_\odot$ (right panel) \cite{LIGOScientific:2018cki}.}
    \label{fig:full_results}
\end{figure*}

In Table  \ref{tab:delta-params} we give the parameters $g_{\rho}$, $g_{\delta}$, and $g_{\omega\rho}g_{\rho}^2$, median and 90\% confidence interval (CI) limits,  resulting from the inference. The corner plot in Figure \ref{fig:corner-NS} shows the three parameters ($g_{\rho}$, $g_{\delta}$, and $g_{\omega\rho}$), three nuclear matter properties connected to the isovector channel ($E_{sym}$, $L$, $K_{sym}$) and some NS properties (the radius of a 1.4 $M_\odot$ and a 2.0 $M_\odot$ NS, $R_{1.4}$ and $R_{2.0}$, the proton fraction $X_p$ at the center of the  maximum mass NS  and the onset density $\rho_{dUrca}$ of the direct Urca processes, the different lines representing  the $1\sigma$ (dotted), $2\sigma$ (dashed), and $3\sigma$ (solid) CI. See also Table \ref{tab:properties} for the median and the 90\% CI of several nuclear matter and neutron star properties. The different  EOS are represented by the colors red  (EOS8), purple (EOS20) and blue (EOS21). The green circle, orange diamond, and the blue star represent the values of the respective quantities before the inclusion of the $\delta$-meson for the EOS8, 20, and 21, respectively, given in Table \ref{tab:params_props}. Note that these properties generally lie inside the $3\sigma$ CI. To allow a complete discussion we have plotted in Fig. \ref{fig:pnm_chieft}  the 90\% credible interval (CI) regions for the pure neutron matter (PNM) pressure (left panel) and energy per neutron of PNM (right panel) as a function of the baryonic density for the EOS8, EOS20, and EOS21 posteriors with the inclusion of the $\delta$-meson. For comparison we also show the  $\chi$EFT N3LO result  for the PNM pressure from Hebeler et al. (2013) \cite{Hebeler2013} in the left panel and the PNM energy per neutron from different  $\chi$EFT calculations as compiled in Ref. \cite{Huth:2021bsp}.
\\

\begin{table*}[]
\centering
\caption{The median and 90\% CI (minimum and maximum) of the model parameters: \(g_\rho\) (the coupling constant for the \(\rho\)-meson), \(g_\delta\) (the coupling constant for the \(\delta\)-meson), and the coupling product \(g_{\omega\rho} \times g_\rho^2\) (representing the non-linear coupling between the \(\omega\) and \(\rho\) mesons scaled by the square of the \(\rho\)-meson coupling, ensuring the third parameter is of the same order of magnitude as the other two). These values are derived from the posterior distributions of the EOS8, EOS20, and EOS21 sets obtained using a Bayesian inference framework that incorporates the \(\delta\)-meson effects.}
\label{tab:delta-params}
\begin{tabular}{cccccccccc}
\hline \hline 
      & \multicolumn{3}{c}{$g_{\rho}$}          & \multicolumn{3}{c}{$g_{\delta}$}     & \multicolumn{3}{c}{$g_{\omega\rho}{g_{\rho}}^{2}$} \\ \cline{2-10} 
      & \multicolumn{3}{c}{90\% CI}             & \multicolumn{3}{c}{90\% CI}          & \multicolumn{3}{c}{90\% CI}                 \\ \cline{2-10} 
EOS   & Median      & Min.        & Max         & Median     & Min.       & Max        & Median       & Min.         & Max           \\ \hline
EOS8  & 13.040591 & 10.312384 & 15.611890 & 1.925159 & 0.154474 & 6.416854 & 7.516723 & 2.688555 & 12.691254   \\
EOS20 & 13.851995 & 10.988632 & 16.032496 & 1.688425 & 0.156531 & 5.340520 & 7.056363 & 3.071680 & 10.616854   \\
EOS21 & 13.163152 & 10.165686 & 15.777018 & 1.875991 & 0.159919 & 6.277619 & 5.231461 & 2.091371 & 8.913382    \\ \hline
\end{tabular}
\end{table*}

We analyze in the following the results shown. The three initial models have quite different isovector properties, with $E_{sym}$ spanning a range of 5 MeV (28.5 - 33.8 MeV), $L$ a range of 10 MeV (from 32 to 42 MeV), and $K_{sym}$ a range of about 125 MeV (from -77 to 56 MeV), see Table \ref{tab:params_props}. In addition, as mentioned already above, the models differ in the predicted maximum mass (from 2.24 to 2.57), in the predicted radius for a 1.4$M_\odot$ star (from 12.59 to 12.95 km), as well as in the expected behavior regarding the opening of direct Urca (dUrca) processes in the NS, with one model not predicting these processes in the NS and the other two models predicting them to occur at 3.5$\rho_0$ and 5$\rho_0$. Figs. \ref{fig:corner-NS} and \ref{fig:pnm_chieft} show that the inclusion of the $\delta$-meson has a large effect on the isospin NMP at saturation,  symmetry energy $E_{sym}$, its slope $L$ and curvature $K_{sym}$ and on the proton fraction $X_p$ and the dUrca onset density ($\rho_{dUrca}$). These effects of the inclusion of the $\delta$-meson on the Bayesian inference of the isovector channel can be summarized as follows: a) the median symmetry energy is uniform, and for the three models the median is $\sim 32$ MeV and the dispersion is of the order of 1.5 MeV; b) a difference between the $L$ distributions is maintained, although the medians are closer. At 3$\sigma$ two of the models reach values above 60 MeV, well above the values of the original models; c) the curvature $K_{sym}$ is the most affected property with three well separated medians at $\sim -10,\, -50$ and -140 MeV, with one of the models allowing values above +50 MeV and another below -200 MeV; d) just by varying the isovector channel through the three couplings it is possible to vary the radius of a 1. 4 $M_\odot$ star by 1 km, from 12.3 to 13.3 km with overlapping distributions, while for the 2.0 $M_\odot$ star there is a variation of 200 to 250 m around the original radii; e) even with the extra freedom given to the isovector channel, the modified EOS8 still almost does not allow dUrca processes inside NS. Table \ref{tab:percentage-durca} shows the percentage of models with the opening of dUrca processes inside NS for the three EOS sets. Unlike EOS20 and 21 sets, which have almost $100\%$ of models with dUrca, EOS8 set has a very low fraction, with less than $1\%$ of models.  Moreover, these few models predict dUrca inside stars close to the maximum mass configuration. The few models that still predict dUrca inside NS have a large $L$ and a quite small coupling $g_\delta$. This last feature indicates that the introduction of the $\delta$-meson can change the behavior of this model concerning the dUrca properties. We will come back to this point when discussing the symmetry energy. For the other two sets, EOS20 and EOS21, the mean $\rho_{dUrca}$  increases slightly, but the new degree of freedom in the Lagrangian allows dUrca to occur already at 3$\rho$ instead of $\sim 4\rho_0$. 

Considering the three couplings $g_\rho$, $g_{\omega\rho}$ and $g_\delta$, we observe a linear correlation between the last two, with a tendency of $g_{\omega\rho}$ to decrease as $g_\delta$ increases, while no correlation is obtained between $g_\rho$ and the other two couplings. The largest values of $g_\delta$ are associated with the smallest values of $g_{\omega\rho}$ and smallest values of $K_{sym}$. Large values of $K_{sym}$ are the result of large values of $g_{\rho}$ and $g_{\omega\rho}$ and intermediate values of $g_\delta$. We conclude that the inclusion of the $\delta$-meson leads to a broadening of the accepted range for the other two couplings, allowing an increase in the accepted range of nuclear matter properties. 

Regarding the PNM properties, we conclude for Fig. \ref{fig:pnm_chieft} that the two different PNM chEFT calculations, one in terms of PNM pressure and the other in terms of PNM energy density, impose different constraints. Recall that the present Bayesian inference was performed imposing the chEFT PNM pressure from \cite{Hebeler2013} at 0.08, 0.12, and 0.16 fm$^{3}$, and all three EOSs satisfy these constraints in this density range. However, EOS20 misses the Hebeler et al. \cite{Hebeler2013} constraints below 0.08fm$^{-3}$ and completely misses the PNM energy density constraints compiled in Huth et al. \cite{Huth:2021bsp}.  Note, however, that EOS8 and EOS21 are fully compatible with chEFT calculations. The inclusion of the $\delta$-meson allows us to explore a parameter range compatible with both sets of ab initio calculations, even for the distribution based on EOS20.

\begin{table}[]
\centering
  \setlength{\tabcolsep}{20pt}
        \renewcommand{\arraystretch}{1.2}
\caption{Percentage of models that present direct Urca process for the three EOS.}
\label{tab:percentage-durca}
\begin{tabular}{cc}
\hline \hline 
EOS    & Percentage with dUrca process (\%) \\ \hline
EOS8  & 0.59                               \\
EOS20 & 99.98                              \\
EOS21 & 99.98                              \\ \hline
\end{tabular}
\end{table}
Figure \ref{fig:full_results} shows the mass \textit{vs.} radius (left panel) and tidal deformability (right panel) plots for the 3 EOS sets. The curves inside the probability distributions resulting from our Bayesian inference represent the three original EOS with no $\delta$-meson. Observational data corresponding to the GW170817 event \cite{LIGOScientific:2018cki} and  the NICER observations of the pulsars PSR J0030 + 0451 \cite{Riley:2019yda,Miller:2019cac}, PSR J0740 + 6620 \cite{Fonseca:2021wxt,Riley:2021pdl,Miller:2021qha} and PSR J0437–4715 \cite{Choudhury:2024xbk} have been included. According to Fig. \ref{fig:full_results}, inserting the $\delta$-meson does not affect much the maximum mass of the NS and the 90\% credible interval (CI) encloses the results without the $\delta$-meson. In addition, the 90\% CI of the probability distributions are compatible with the observational data available.  Note that the distribution obtained is compatible with the recent NICER observation of the pulsar PSR J1231-1411 with a mass 1.04$\scriptsize^{+0.05}_{-0.03}\, M_\odot$ and a radius 12.6$\pm$0.3 km or 13.5$\scriptsize^{+0.03}_{-0.05}$ at a 68\% CI depending on the prior used, since we get at 90\% CI 11.84$<R_{1.0}<$12.79~km (EOS8), 12.11$<R_{1.0}<$12.82~km (EOS20) and  12.28 $<R_{1.0}<$13.05~km (EOS21).

Regarding the tidal deformability (Fig. \ref{fig:full_results} right panel), we verify that the main effect of introducing the $\delta$-meson is to allow the access of a $M\Lambda$ region with smaller  $\Lambda$ values, with the distribution of each set lying practically below the reference EOS. The blue bar identifies the constraint obtained in \cite{LIGOScientific:2018cki} for the effective tidal deformability, $\tilde\Lambda<720$ considering the binary mass ratio $q=m_2/m_1=1$, which corresponds to $M=1.36M_\odot$.

The isospin channel affects strongly the radius of medium and low mass stars. This effect of the isovector channel has been frequently discussed in studies that did not include the $\delta$-meson, see \cite{Horowitz:2000xj,Carriere:2002bx,Cavagnoli:2011ft}.  In \cite{Reed:2023cap}, the $\delta$-meson was introduced and it was shown that it also affects the NS radii. The EOS8, EOS20 and EOS21 low and medium mass radii are determined both by the stiffness of the EOS as defined by $K_0$, $Q_0$ and $Z_0$ and by their symmetry energy properties at saturation, in particular, the $L$ value. Introducing the $\delta$-meson the range spanned by the radius of 1.0 $M_\odot$ stars has increased from $\sim 200$m  to $\sim$ 1200 m. In particular, EOS8 has a very soft symmetry energy, which compensates the large incompressibility, and presents the smallest radii for low and medium mass stars.

\begin{figure}[h]
    \centering
    \includegraphics[width=1.\linewidth]{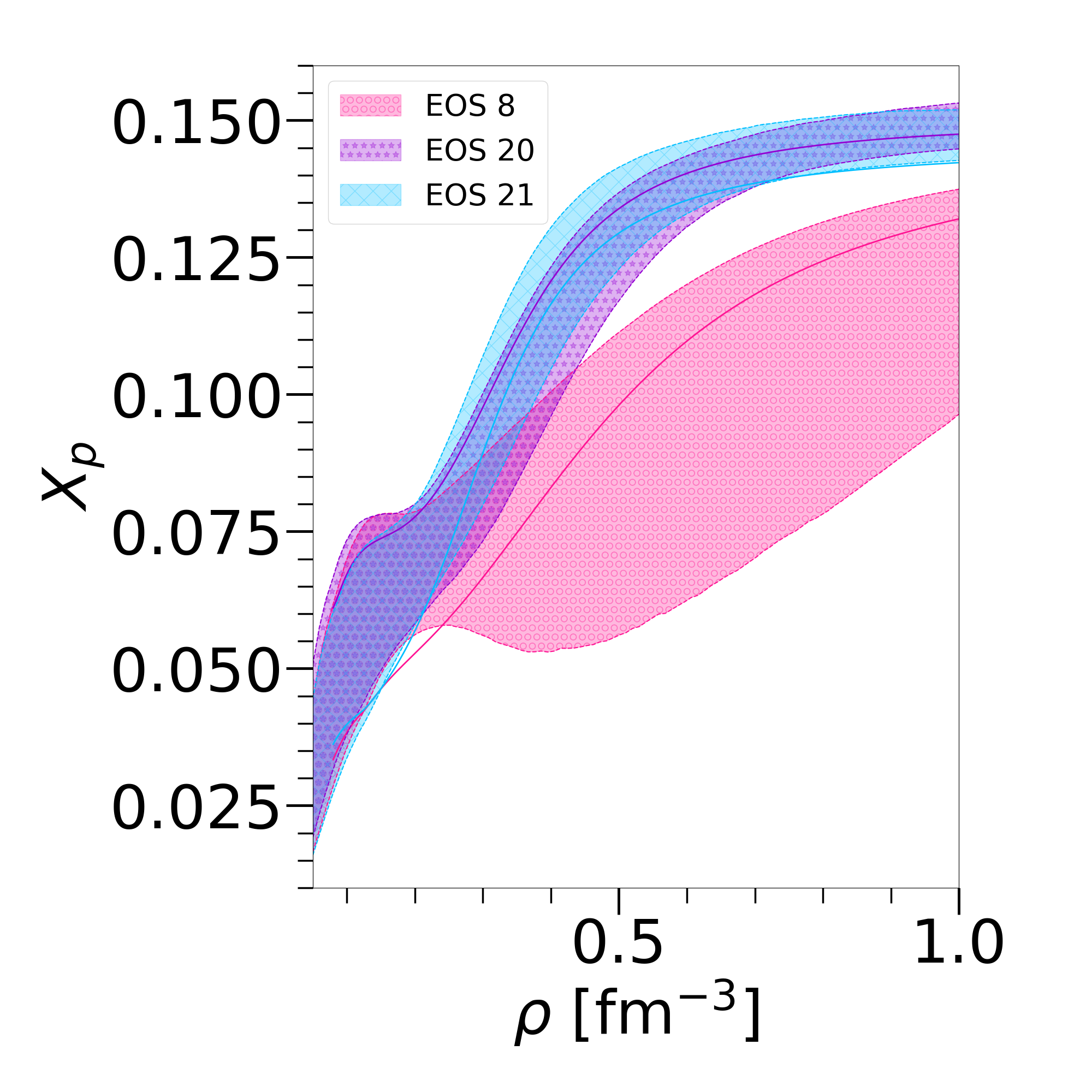}\\
    \includegraphics[width=1.\linewidth]{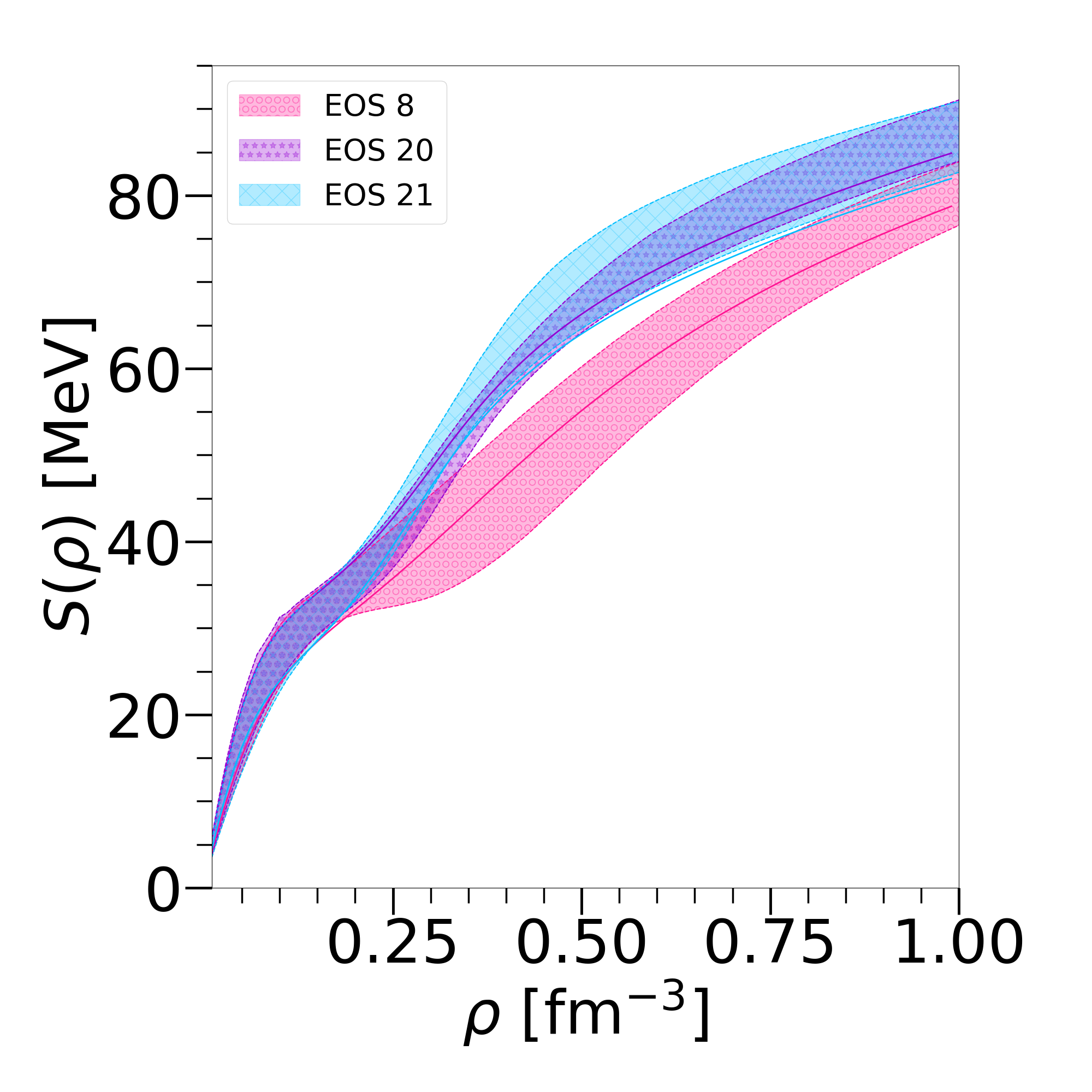}
    \caption{The region representing the 90\% confidence interval for  (bottom) the symmetry energy $S(\rho)$ as a function of baryon density $\rho$ and (top) the proton fraction in $\beta$-equilibrium matter for the considered EOS sets 8, 20, and 21. The solid line shows the plot in the corresponding quantity on each panel excluding the $\delta$-meson. }    
    \label{fig:proton-fraction}
\end{figure}

\begin{figure}[h]
    \centering
    \includegraphics[width=1.\linewidth]{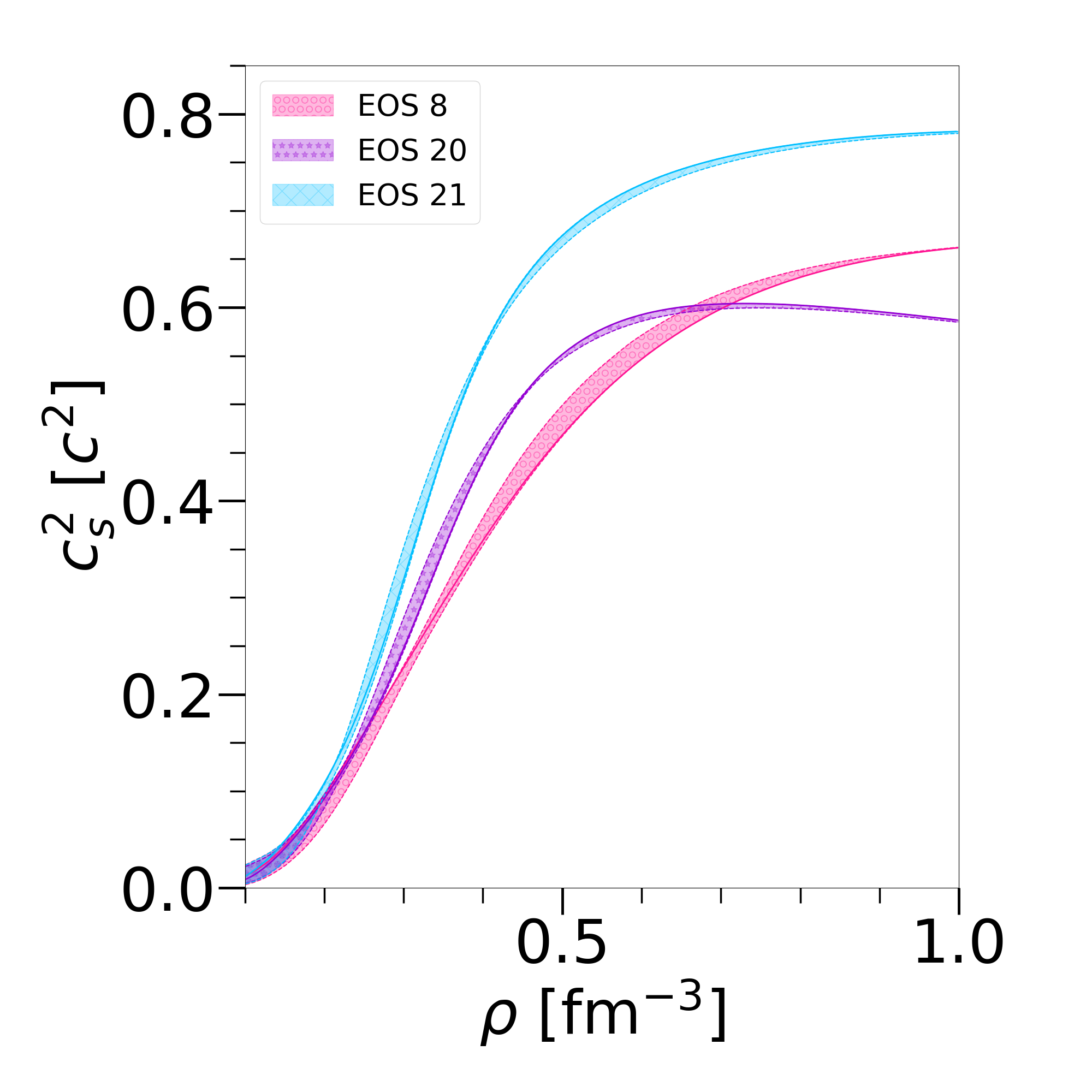}
    \caption{The region representing the 90\% confidence interval for the sound speed squared in $\beta$-equilibrium matter as a function of the baryonic density $\rho$ for the three EOS sets. The solid lines represent the EOS8, 20 and 21 without the $\delta$-meson.}
    \label{fig:sound-speed-vs-dens}
\end{figure}

\begin{figure}
    \centering
    \includegraphics[width=1\linewidth]{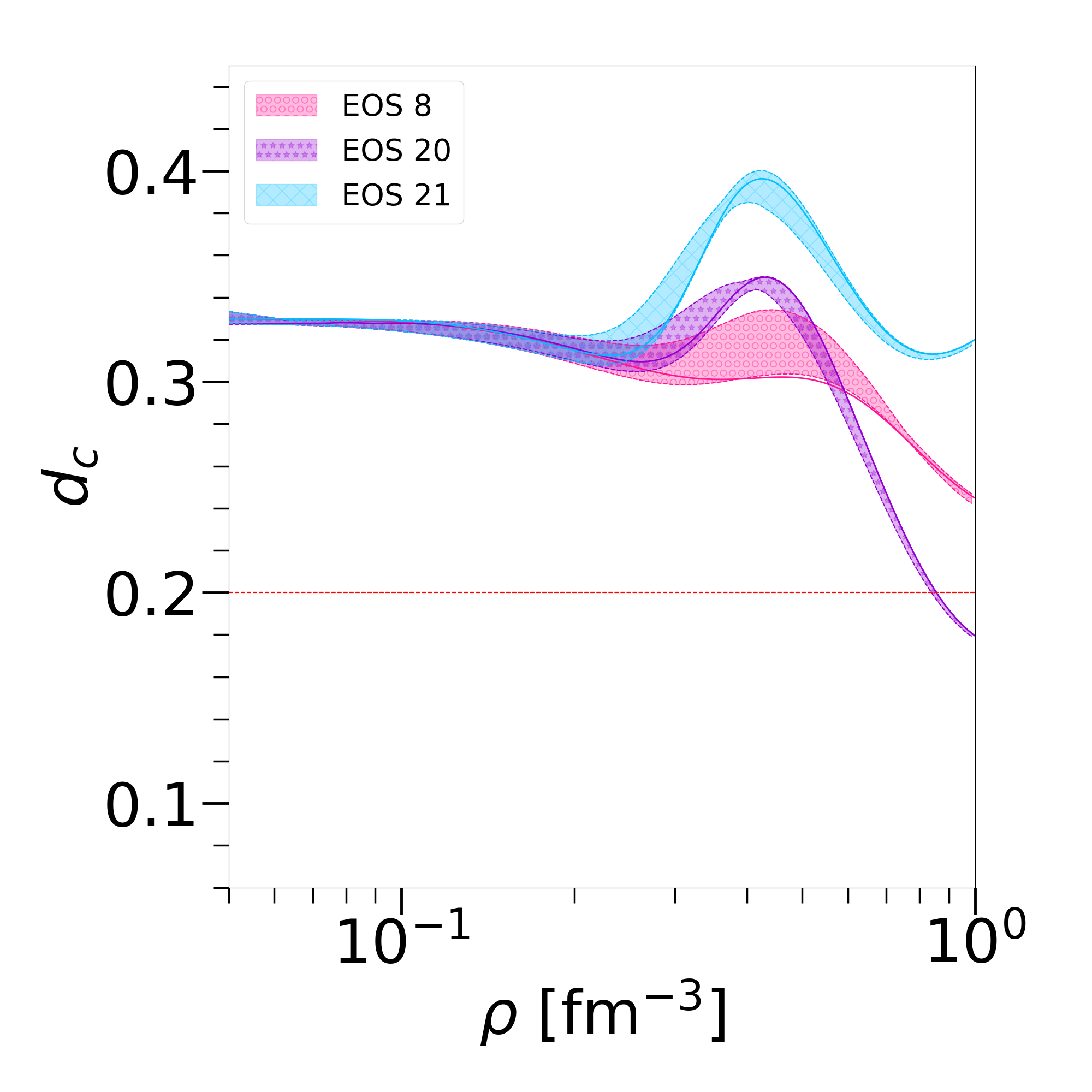}
    \caption{The 90\% confidence intervals for $d_c$ are displayed as a function of $\rho$ for EOS8, EOS20, and EOS21. Here, $d_c=\sqrt{\Delta^2 + {\Delta}'^2}$, where $\Delta' = c_s^2 \left(\frac{1}{\gamma}-1\right)$ represents the logarithmic derivative of $\Delta = \frac{1}{3}-\frac{P}{\epsilon}$ with respect to the energy density, which tends to zero in the conformal limit \cite{Annala:2023cwx}. The solid lines represent the values for the models without the inclusion of the $\delta$-meson.}
    \label{fig:dc}
\end{figure}

\begin{figure}[h]
    \centering    
    \includegraphics[width=1.\linewidth]{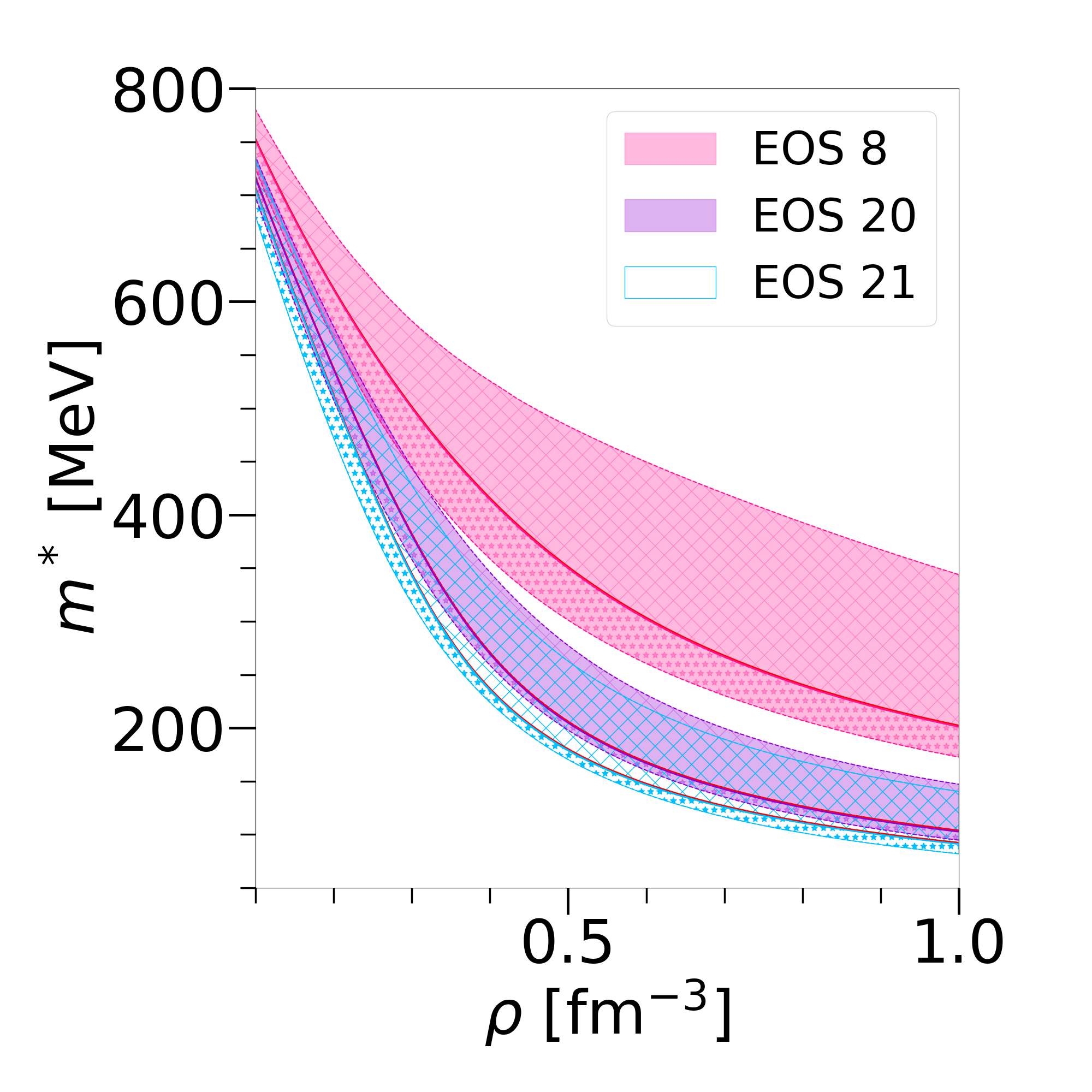}
    \caption{The regions representing the 90\% confidence interval of the effective Dirac masses as a function of density for the three EOS sets. The solid lines represent the EOS without the $\delta$-meson, while the hatched areas represent the EOS sets, for the neutron ($\star$) and proton (X) effective masses, respectively.}
    \label{fig:meff-vs-dens}
\end{figure}

Quantities that measure the isovector channel of the EOS are the symmetry energy and the proton fraction of $\beta$-equilibrium matter, both represented in Fig. \ref{fig:proton-fraction}, respectively, in the bottom and top panels. In particular, the proton fraction defines the onset of the nucleonic direct Urca processes, and, therefore, the cooling evolution of the NS. In both plots the bands represent the probability distributions within the 90\% CI and the original EOS are represented by full lines. The effects of changing the three parameters defining the isovector channel are particularly strong at low densities when the original values may be outside the 90\% CI. Above 0.25 fm$^{-3}$  the effects are both in the direction of stiffening the symmetry energy and, therefore, increasing the proton fraction inside neutron stars, or softening it with the opposite effect on the proton fraction.

The EOS8 90\% CI proton fraction distribution is the broadest, indicating a greater variability of the proton fraction for this EOS than for the others. This is the EOS that accepts a wider range for the $g_{\omega\rho}$ coupling. The EOS8 set shows the softest behavior of the symmetry energy above the saturation density, and this implies that the proton fraction inside the NS remains below 0.14 up to a baryonic density above 1 fm$^{-3}$. As discussed earlier, a direct consequence is the almost total absence of dUrca processes inside stars described by EOS of this set.
At high densities, $\rho \gtrsim 0.6 $ fm$^{-3}$, EOS20 and 21 show similar values of the proton fraction, which can go above 0.15 for $\rho<1$fm$^{-3}$ and thus higher than that for EOS8 in the same range. For these two models the median for the dUrca onset occurs around 4$\rho_0$. 

It is seen that the symmetry energy  clearly follows the behavior of the proton fraction, as expected. All sets show a similar behavior below the saturation density, and above this density the EOS8 set presents the softer symmetry energy, which explains the smallest proton fractions at high densities: the asymmetry proton-neutron does not cost as much energy as for the other two sets.

The sound speed $c_s^2$ at the center of the maximum mass star, $M_{max}$, is shown in Fig. \ref{fig:sound-speed-vs-dens} as a function of density $\rho$. The isospin channel has very little effect on the behavior of the sound speed, which is essentially determined by the isoscalar channel, and hence the distributions are narrow bands along the EOS8, EOS20 and EOS21 curves. The relative behavior of the three distributions is easily understood: i) EOS20 shows a maximum at $\sim$ 0.5 fm$^{-3}$ due to the large value of the coupling $\xi$, 10 (2) times larger than the corresponding value of EOS21 (EOS8). It has been discussed in \cite{Mueller1996} that under these conditions the square of the speed of sound goes to 1/3 at large densities; ii) the EOS21 is particularly hard at large densities, as expected for the large values of the parameters $Q_0$ and $Z_0$. This behavior also reflects the very small $\xi$ coupling, so that the behavior is defined by the terms in the pressure and energy density that are proportional to the density squared.

Fig. \ref{fig:dc} shows the $d_c$ a quantity that was introduce to measure of conformality \cite{Annala:2023cwx},   $d_c=\sqrt{\Delta^2 + {\Delta}'^2}$, where $\Delta' = c_s^2 \left(\frac{1}{\gamma}-1\right)$ represents the logarithmic derivative of the normalized trace anomaly introduced in \cite{Fujimoto:2022ohj} $\Delta = \frac{1}{3}-\frac{P}{\epsilon}$ with respect to the energy density, which tends to zero in the conformal limit. Similar to the speed of sound, this quantity is only slightly affected by the isovector interaction, and the main behavior of each band is determined by the behavior of the three reference EOSs chosen as the basis for the present study. In \cite{Annala:2023cwx}, the value 0.2 was considered to be a value below which the presence of deconfined matter is expected. It has already been shown that, depending on the high-density behavior of a hadronic EOS, these values can also be satisfied by pure hadronic matter. \cite{Providencia:2023rxc,Malik:2023mnx,Takatsy:2023xzf,Malik:2024qjw}. The three reference EOS, represent precisely three quite different behaviors of $d_c$: EOS20 with a soft behavior at high densities as indicated by the speed of sound squared, takes values below 0.2 above densities of the order of six times saturation density, while EOS21 shows no tendency to come close to 0.2.

The peculiarity of the inclusion of the $\delta$-meson is the introduction of a different density dependence of the proton and neutron Dirac masses in asymmetric matter. The scalar Dirac mass contributes to the spin-orbit potential. Note that the Dirac mass is different from the Landau mass which defines the density of states, see \cite{Chamseddine:2023voh} for a recent discussion. In Fig. \ref{fig:meff-vs-dens}, the 90\% CI probability distributions of the effective masses of the nucleons (neutrons, denoted by "$\star$", and protons, denoted by "X") are plotted as a function of density.  The solid curves representing EOS8, 20, and 21 lie exactly on the boundary between the probability distributions of their respective EOS sets. The effective neutron and proton masses of $\beta$-equilibrium matter for a given density decrease and increase, respectively, away from no $\delta$-meson effective mass. 

In Table \ref{tab:properties}, we collect some NMP and NS properties, in particular, the median and the 90\% CI values. Some of these values have already been discussed. It is worth pointing out that the tidal deformability of a 1.4$M_\odot$ can suffer a quite large reduction with an adequate choice of the parameters.  Also the onset of the dUrca processes may vary in a quite large range of densities for sets EOS20 and EOS21 with the 90\% CI extremes differing $\gtrsim 0.2$~fm$^{-3}$. This range is however, below 0.1~fm$^{-3}$ for EOS8,  and in all cases dUrca is possible, its  onset occurs inside NS with a mass close to the maximum mass configuration.

\begin{table*}[]
\caption{\label{tab:properties} Some properties values for EOS8, 20, and 21 after the inclusion of the $\delta$-meson as maximum mass star $M_{max}$, radii ($R$) and tidal deformability ($\Lambda$) at $M=1.4M_\odot$, $M=1.6M_\odot$, $M=1.8M_\odot$, and at $M=2M_\odot$, proton fraction ($X_p$) and symmetry energy ($\mathcal{S}$) for 1$\sigma$ (0.32), 2$\sigma$ (0.64), and 3$\sigma$ (0.96) intervals.}
\setlength{\tabcolsep}{10.pt}
\renewcommand{\arraystretch}{1.2}
\begin{tabular}{clccccccccc}
\hline \hline 
\multirow{3}{*}{Quantity} & \multirow{3}{*}{Units} & \multicolumn{3}{c}{EOS8}                             & \multicolumn{3}{c}{EOS20}                            & \multicolumn{3}{c}{EOS21}                            \\ \cline{3-11} 
                          &                        & \multirow{2}{*}{median} & \multicolumn{2}{c}{90\% CI} & \multirow{2}{*}{median} & \multicolumn{2}{c}{90\% CI} & \multirow{2}{*}{median} & \multicolumn{2}{c}{90\% CI} \\ \cline{4-5} \cline{7-8} \cline{10-11} 
                          &                        &                         & min          & max          &                         & min          & max          &                         & min          & max          \\ \hline

$M_{\rm max}$                    & $M_\odot$              & 2.229                   & 2.222        & 2.246        & 2.344                   & 2.334        & 2.361        & 2.561                   & 2.554        & 2.573        \\
$R_{max}$                    & \multirow{8}{*}{km}    &11.21                    &11.11         &12.41         &11.68                    &11.62         &11.77         &12.14                    &12.08         &12.24         \\
$R_{1.0}$              &      &     12.25     &11.84                   &12.79         &12.38         &12.11                    &12.82        &12.57                    &12.28         &13.05         \\
$R_{1.2}$                    &     &12.38                    &11.99         &12.80         &12.50                    &12.29         &12.87        &12.74                    &12.50         &13.14         \\
$R_{1.4}$                    &     & 12.44                   & 12.10        & 12.80        & 12.61                   & 12.43        & 12.91        & 12.89                   & 12.69        & 13.22        \\
$R_{1.6}$                    &                        & 12.45                   & 12.15        & 12.74        & 12.67                   & 12.53        & 12.92        & 13.01                   & 12.84        & 13.29        \\
$R_{1.8}$                    &                        & 12.40                   & 12.14        & 12.62        & 12.69                   & 12.58        & 12.89        & 13.09                   & 12.95        & 13.33        \\
$R_{2.0}$                    &                        & 12.22                   & 12.03        & 12.38        & 12.63                   & 12.54        & 12.80        & 13.12                   & 13.01        & 13.32        \\
$R_{2.08}$                    &                        & 12.08                   &11.92         &12.22         &12.57                    &12.48         &12.73         &13.12                    &13.01         &13.30         \\
$\Lambda_{1.4}$                  & \multirow{4}{*}{...}   & 475                     & 405          & 517          & 524                     & 502          & 556          & 615                     & 590          & 662          \\
$\Lambda_{1.6}$                  &                        & 207                     & 181          & 221          & 238                     & 230          & 250          & 291                     & 281          & 310          \\
$\Lambda_{1.8}$                  &                        & 90                      & 81           & 95           & 111                     & 107          & 116          & 143                     & 138          & 151          \\
$\Lambda_{2.0}$                  &                        & 37                      & 34           & 39           & 50                      & 49           & 53           & 70                      & 68           & 75           \\
{$X_p (0.32)$}                & \multirow{3}{*}{...}   & 0.077                   & 0.055        & 0.091        & 0.096                   & 0.077        & 0.106        & 0.102                   & 0.085        & 0.113        \\
{$X_p (0.64)$}                &                        & 0.113                   & 0.064        & 0.123        & 0.140                   & 0.134        & 0.146        & 0.140                   & 0.135        & 0.148        \\
{$X_p (0.96)$}                &                        & 0.130                   & 0.092        & 0.137        & 0.147                   & 0.144        & 0.153        & 0.144                   & 0.142        & 0.153        \\
{$\mathcal{S}(0.32)$}                & \multirow{4}{*}{MeV}   & 42.80                   & 34.23        & 46.95        & 49.13                   & 45.71        & 51.83        & 51.25                   & 48.74        & 54.78        \\
{$\mathcal{S}(0.64)$}                 &                        & 64.49                   & 57.48        & 68.94        & 73.34                   & 71.57        & 78.33        & 72.82                   & 71.18        & 81.52        \\
{$\mathcal{S}(0.96)$}                 &                        & 80.73                   & 77.80        & 85.60        & 86.14                   & 84.52        & 92.60        & 84.18                   & 83.01        & 92.10        \\
{$\mathcal{S}$}                   &              & 32.03                   & 29.20        & 34.42        & 31.78                   & 28.97        & 34.44        & 32.00                   & 29.23        & 34.63                      \\
$L$                    & MeV              & 32.1                   & 14.5        & 56.0        & 32.9                   & 18.5        & 50.2        & 40.7                   & 23.6        & 60.3        \\
$K_{sym}$                    & MeV              & -138                   & -250        & -70        & -49                   & -124        & 10        & -10                   & -70        & 55        \\$\rho_{dUrca}$                    & fm$^{-3}$              & 0.959                   & 0.904        & 0.988        & 0.608                   & 0.521        & 0.715        & 0.588                   & 0.464        & 0.730        \\$M_{dUrca}$                    & $M_\odot$              &2.219                    &2.217         &2.222         &2.196                    &1.968         &2.330         &2.446                    &2.045         &2.566         \\
$c^{2}_{s}$                    & $c^{2}$              &0.6589                    &0.6582       &0.6597         &0.5966                    &0.5948         &0.5969         &0.7679                    &0.7628         &0.7685         \\
$\rho_{c}(max)$                    & fm$^{-3}$              &0.966                    &0.962         &0.971         &0.882                    &0.870         &0.886         &0.791                    &0.782         &0.794         \\\hline
\end{tabular}
\end{table*}

\begin{table}[]
\centering
\caption{The median and 68\% CI min and max of NS radii values at 1.4$M_\odot$ for the three EOS sets.}
\label{tab:r14_68CI}
\setlength{\tabcolsep}{18.pt}
\renewcommand{\arraystretch}{1.1}
\begin{tabular}{cccc}
\hline \hline 
\multirow{3}{*}{EoS} & \multicolumn{3}{c}{$R_{1.4}$(km)} \\
                     & \multicolumn{3}{c}{68\% CI} \\
                     & Median   & Min     & Max    \\ \hline
EoS 8                & 12.44    & 12.26   & 12.64  \\
EoS 20               & 12.61    & 12.49   & 12.77  \\
EoS 21               & 12.89    & 12.76   & 13.09  \\ \hline
\end{tabular}
\end{table}

\section{Conclusions}
The present work is an exploratory study aimed at understanding the impact of including the scalar-isovector \(\delta\)-meson on the description of nuclear matter and neutron star properties within the Relativistic Mean-Field (RMF) framework. This $\delta$ meson has a direct effect on the isovector channel of nuclear matter, affecting the density dependence of the symmetry energy. It also gives rise to a splitting of the Dirac mass for protons and neutrons in asymmetric matter.  Starting from three EOS without the $\delta$-meson with acceptable properties and predicting different maximum star masses, we have performed a Bayesian inference calculation taking each EOS independently to constrain the three couplings directly related to the symmetry energy. Some minimal nuclear matter properties at saturation, chEFT constraints on neutron matter and the condition of describing two solar mass neutron stars have been imposed. 

The three sets of EOS obtained from the reference EOS match the available NS observations, such as those from the GW170817 event \cite{LIGOScientific:2018cki}, NICER observations of the pulsars PSR J0030 + 0451 \cite{Riley:2019yda,Miller:2019cac}, PSR J0740 + 6620 \cite{Fonseca:2021wxt,Riley:2021pdl,Miller:2021qha} and PSR J1231-1411 \cite{Salmi:2024bss}. However,  there is some tension with the radius values obtained for PSR J0437–4715 \cite{Choudhury:2024xbk} with a mass $\sim 1.4M_\odot$ and a radius $10.73<R<12.31$ km at 68\% CI; see Table \ref{tab:r14_68CI}. The maximum mass of the NS was shown to be not much affected by the inclusion of the $\delta$-meson. Properties such as the speed of sound squared and the quantity $d_c$ associated with the trace anomaly also do not depend much on the isovector channel.

The three couplings that govern the isospin channel in this model have an important effect on: i) the radius of low and medium mass stars; ii) the proton fraction inside neutron stars in $\beta$-equilibrium; iii) some properties of nuclear matter connected to the symmetry energy, in particular,  allowing the symmetry energy curvature to take values close to or above zero, predicting larger values for the symmetry energy at equilibrium, which can be as high as 35 MeV within a 90\% CI; and allowing the symmetry energy slope to vary between 15 and 60 MeV. These properties affect mainly the proton fraction, which at low densities suffers a quite large increase. In addition, a split between the proton and neutron Dirac masses is observed. 

The inclusion of this extra meson in the model has allowed one to enlarge the allowed phase space spanned both for the EOS and for the mass-radius compatible with accepted properties. In particular, the $\delta$-meson has brought more flexibility to the RMF description in the isospin channel. Future works should explore more than the three parameters studied in the present paper ($g_{\rho}$, $g_{\delta}$, and $g_{\omega\rho}{g_{\rho}}^{2}$), and in particular extend to the complete set of model parameters. In fact, given the current uncertainties, neutron star mass and radius measurements alone cannot fully disentangle the relative contributions of the isoscalar and isovector sectors of the nuclear equation of state \cite{Imam2021,Tovar2021,Mondal2021,Essick2021}. However, as future gravitational wave and X-ray observations, combined with more precise nuclear experiments, further refine our understanding of isovector properties, incorporating the $\delta$-meson will likely prove essential for achieving a comprehensive and realistic description of nuclear matter and neutron star interiors.

\section{Acknowledgements} 
This work was partially supported by national funds from FCT (Fundação para a Ciência e a Tecnologia, I.P, Portugal) under projects UIDB/04564/2020 and UIDP/04564/2020, with DOI identifiers 10.54499/UIDB/04564/2020 and 10.54499/UIDP/04564/2020, respectively, and the project 2022.06460.PTDC with the associated DOI identifier 10.54499/2022.06460.PTDC. The authors acknowledge the Laboratory for Advanced Computing at the University of Coimbra for providing {HPC} resources that have contributed to the research results reported in this article, URL: \hyperlink{https://www.uc.pt/lca}{https://www.uc.pt/lca}. 

\bibliographystyle{apsrev4-2}
\bibliography{biblio}
\end{document}